\documentclass[preprint,showpacs,preprintnumbers,amsmath,amssymb]{revtex4}
\usepackage{graphicx}
\usepackage{dcolumn}
\usepackage{bm}
\usepackage{amssymb}
\usepackage{amsmath}

\begin{document}
\title{Characterizing the insulator adjacent to the superconductor in Bi$_2$Sr$_{2-x}$La$_{x}$CuO$_{6+\delta}$ ($x=0.3$)}

\author{L. Fruchter, Z. Z. Li and H. Raffy
}                     
\affiliation{Laboratoire de Physique des Solides, C.N.R.S.
Universit\'{e} Paris-Sud, 91405 Orsay cedex, France}
\date{Received: date / Revised version: date}
%
\begin{abstract}{
The time-averaged and low--frequency noise transport properties were investigated in the vicinity of the superconductor--insulator transition for a Bi$_2$Sr$_{2-x}$La$_{x}$CuO$_{6+\delta}$ ($x=0.3$) thin film. The results are consistent with a superconductor (metal) embedded in a strong insulator, the latter showing two-dimensional variable range hopping properties. The weak insulator behavior -- if any -- is attributed to the metallic inclusions only.}
\end{abstract}

\pacs{71.30.+h,72.70.+m,74.72.Hs}

\maketitle
An approach for the understanding of superconductivity in high-temperature superconductor is to start from the insulator parent compound and to increase the carrier concentration to obtain the underdoped superconductor. Within such a perspective, the physics of high-T$_c$ superconductivity is that of strongly correlated electrons, or holes, of the Mott insulator, for which charge repulsion plays the essential role (although the contribution of antiferromagnetism to the properties of the lightly doped material is still under debate\cite{comanac2007}). As very early pointed out, there are strong similarities between the high-T$_c$ superconductors and doped two-dimensional semiconductors\cite{edwards1995,edwards1998}. Indeed, high-T$_c$ cuprates may be viewed as weakly compensated semiconductors, the effective Bohr radius for the excess charge being of the order of a few \AA: for hole-doped cuprates, the role of acceptors is played by the chemical substituant or doping oxygen, and the one of donors by the copper in the CuO$_2$ plane - the latter greatly outnumbering the former. There are important differences also, and the role of fermions may be played in this case by bosons that account for some real-space pairing before condensation occurs\cite{edwards1998}. The metal emerging from the insulator is also peculiar and exhibits a pseudo-gap (which is not a Coulomb gap). There are discussions on how the transition from the antiferromagnetic insulator to the superconductor (alt. to the metal at finite temperature) exactly occurs. First, according to Rullier-Albenque et al\cite{rullier2008}, the antiferromagnet insulator turns directly into the `normal' metal  in the absence of disorder only, otherwise an intermediate spin-glass--like regime is found. Indeed, heavily underdoped Bi$_2$Sr$_2$CuO$_{6+\delta}$ exhibit negative magnetoresistance reminiscent of some spin-flip scattering\cite{fruchter2006}. Then, it was recently proposed\cite{oh2006} that the reentrant resistive behavior observed in almost all underdoped cuprates, as $T$ is lowered, may be accounted for by a phase separation mechanism between an insulating and a superconducting phase. Direct -- although surface -- observations for Bi-based cuprates show nanoscale inhomogeneity\cite{cren2000,kugler2001,davis2005,mashima2006}, which could be either intrinsic to these cuprates for which there is a local doping effect, or originate from chemical inhomogeneities or substitutions due to the elaboration process. Such a reentrant behavior is also predicted in the case of a conventional granular superconductor\cite{strelniker2007}.

Charge transport in semiconductors and at the metal-insulator transition is a well documented topic. In particular, the incidence on transport noise  of disorder randomly distributed in space -- as it is the case for cuprates -- was investigated both theoretically and experimentally\cite{shklovskii1980,kogan1980,kogan1998,kozub1996,pokrovskii2001,shklovskii2003}. These studies provide information on the topological distribution of charges and their possible interactions. A lot of attention has been paid to the conventional transport properties in cuprates, but much less attention was given to transport noise properties. While one might expect that transport noise in the insulating regime bears strong resemblance with that for semiconductors, as stated above, an intermediate situation where the superconductor possibly coexists with the insulator could be quite specific. Indeed, charge fluctuations in the insulator are thought to be governed by the simple quantum tunneling mechanism between localized states, whereas charge transfer between the insulator and the superconductor is a more complex problem, which cannot simply be described using the traditional Andreev reflection description, due to the lack of any well defined charge momentum for localized states\cite{kozub2006}.

We studied a Bi$_2$Sr$_{2-x}$La$_x$CuO$_{6+\delta}$ thin film, with $x=0.3$. The effect of the lanthanum substitution is to remove charges from the CuO$_2$ planes, yielding states less doped than for the pure compound. As a result, full doping, using pure oxygen gas annealing, yields only slightly overdoped states, while the full doping range may be obtained for the un-substituted material ($x=0$). An additional effect is to reduce the CuO$_2$ plane modulation which, for the pure material, is thought to contribute to the reduction of the maximum superconducting temperature for this compound\cite{li2005} ($T_c \approx$ 20 K, x = 0; $T_c \approx$ 30 K, x = 0.3). As a result, one might expect that the La--substituted, underdoped, material is less disordered than the pure one. Different doping states were attained by first annealing the samples in pure oxygen flow at 420$^\circ$C to obtain maximum doping, and then annealing under vacuum in the temperature range 220$^\circ$C -- 260$^\circ$C to successively obtain less doped states. The film was $1950$~\AA{} thick, epitaxially grown on a SrTiO$_3$ substrate. In order to minimize the contribution from the current source noise and maximize the signal, the film was patterned in the Wheatstone bridge geometry, with four strips $17$ x $370\,\mu$m made from electron beam lithography. Gold contacts were deposited using rf-sputtering. The bridge unbalance was 3 \%, likely arising from geometrical imperfections. Noise measurements were performed nulling this unbalance, using an external decade resistor. The temperature of the sample was carefully stabilized, in order to limit the contribution of the temperature fluctuations to the low frequency--voltage noise : monitoring of the sample temperature yielded a typical $0.5$ mK standard deviation from the setpoint. The low frequency noise spectrum was measured using the demodulation technique\cite{scofield1987,fruchter2006}. In order to get rid of spurious voltage spikes which are sometimes oberved in a noisy laboratory environment (typically a few events per hour, in our case), the raw voltage signal, after anti-aliasing filtering by the spectrum analyzer, was recorded and subsequently filtered and Fourier transformed to get the voltage noise power spectrum is obtained, $S_v(f)$, from which the normalized spectral density, $S=S_v(f)/V^2$ ($V$ is the voltage applied to the sample).

\begin{figure}
\resizebox{0.75\columnwidth}{!}{%
\includegraphics{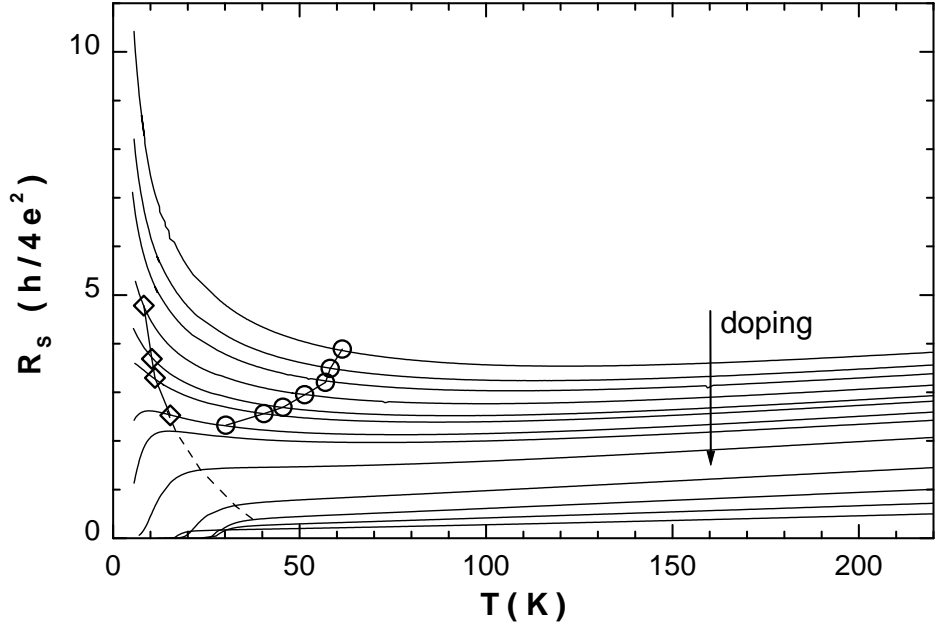}}
\caption{Sheet resistance at various doping levels. Circles and diamonds delimit the VRH regime in Fig.\,\ref{representations}a.}\label{resistance}
\end{figure}

\begin{figure}
\resizebox{0.75\columnwidth}{!}{%
\includegraphics{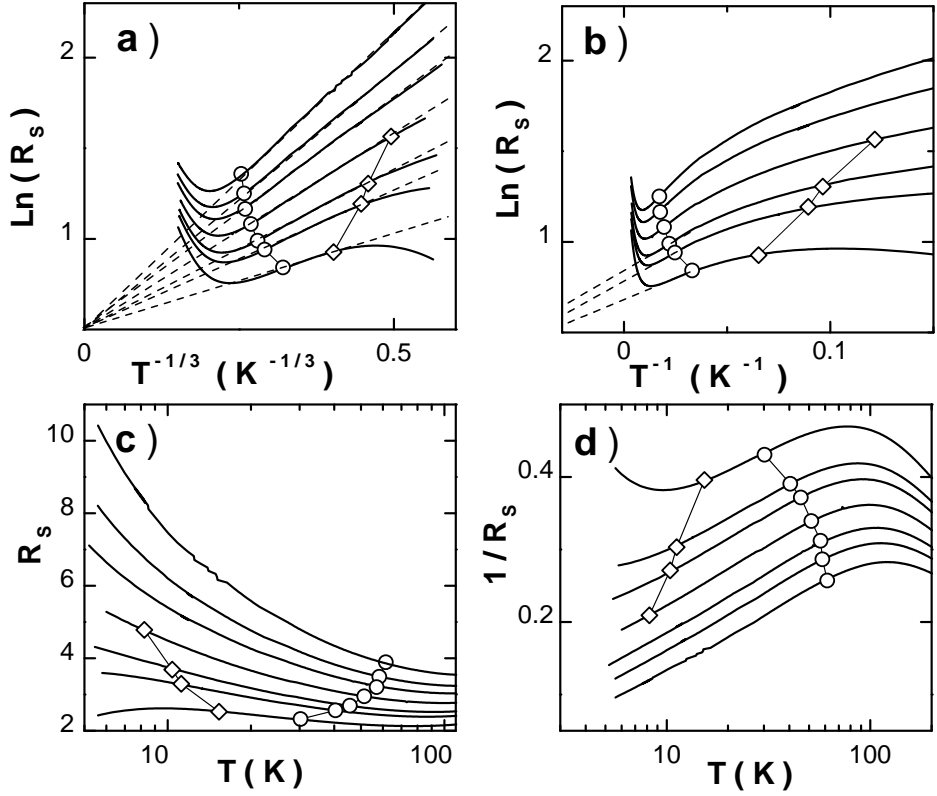}}
\caption{Different representations for the data in Fig.\,\ref{resistance} (starting from the less doped state), in units for a linear representation for : a) 2D variable hopping (dashed lines; b) thermally activated behavior for nearest neighbor hopping; c) Kondo scattering and weak localization; d) alternative representation for weak localization. Circles and diamonds -- also plotted in Fig.~\ref{resistance} -- indicate the linear behavior in Fig.~a}\label{representations}
\end{figure}

The time-averaged two-dimensional resistance per square and per CuO$_2$ plane (sheet resistance, $R_s$) for several doping states is displayed in Fig.\,\ref{resistance}, in units of the universal quantum resistance for pairs, $R_Q = h / 4 e^2 = 6.5$~k$\Omega$. In a way which is generic for all cuprates\cite{oh2006}, three regimes are observed: for the lower doping states it continuously increases for decreasing temperature; in a short range of doping, $R(T)$ increases before superconductivity finally occurs (re-entrant regime), and, for the higher doping, the resistivity monotonically decreases with temperature. Superconductivity is here observed for resistivity well above the critical value $R_Q$. The low temperature resistivity upturn has been the subject of many experimental investigations, but the situation is not yet fully clarified. Fig.\,\ref{representations} displays the main representations and classical justifications that may be found in the literature. Kondo--like scattering ($R \sim \log{(1/T)}$) was found to be a convincing explanation for both large resistivity increase with decreasing temperature in high magnetic field and negative, \textit{isotropic} magnetoresistance, although the expected saturation of the effect by the magnetic field was rarely found at low temperature\cite{rullier2008}. 2D weak localization fails to account for both isotropic M.R. and saturation, but it was nevertheless found to account for both resistivity and orbital magnetoresistance in several cases\cite{jing1991,fournier2000}. In the present case, the Kondo--like behavior conveniently fits the data only for some particular doping level, while larger and smaller doping exhibit downwards and upwards deviations from the expected linear behavior, in the $\log$ scale in Fig.\,\ref{representations}c (subtraction of some background obtained from the high--temperature resistivity data did not improve this). 2D weak localization behavior ($1/R \sim \log{(T)}$, conductivity being the quantity actually affected by the logarithmic correction) is compatible with the data for the lower doping (Fig.\,\ref{representations}d) but, as shown in Fig.\,\ref{representations}a, over some range of temperature and doping, resistivity may also be described using the two dimensional variable range hoping (VRH) law, $R(T) = R_0 \exp{(-T/T_0)}^{1/3}$, using the \textit{same} prefactor $R_0 = 1.65 \, R_Q$ for all curves. The three dimensional VRH law (exponent $1/4$) was found to fit the data over smaller temperature ranges, and free parameter fits of the less doped states yielded exponents in the range $0.30 - 0.34$. At low temperature and in some doping range, the data clearly falls \textit{below} the VRH linear representation (diamonds in Fig.\,\ref{resistance} and\,\ref{representations}a). Similarly, at high temperature, the data falls \textit{above} the VRH linear representation (circles in Fig.\,\ref{resistance} and \,\ref{representations}a). We attribute the low temperature deviation to a superconducting contribution, while the high temperature one may be tentatively described -- over a narrow temperature range -- by a thermally activated law, usually attributed to the nearest neighbor hopping regime (NNH) (Fig.\,\ref{representations}b). Clearly, although the domains where superconductivity and VRH manifest themselves appear as being adjacent for the most underdoped states in Fig.\,\ref{resistance}, it becomes more difficult to assess which transport mechanism dominates as doping increases and one enters the re-entrant regime, as no modelization is then valid over a broad temperature range.

\begin{figure}
\resizebox{0.75\columnwidth}{!}{%
\includegraphics{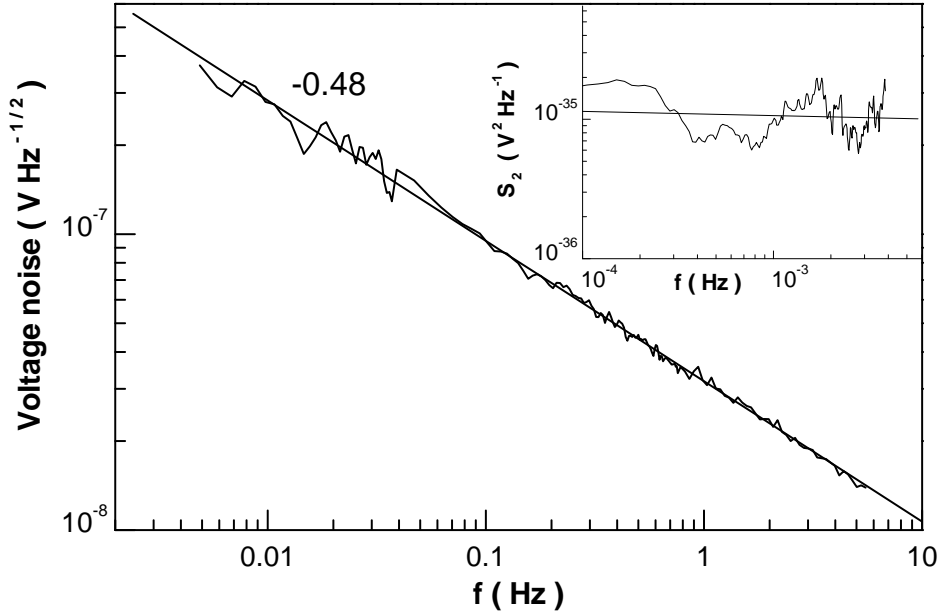}}
\caption{Voltage noise spectrum for sample with $R(300 K) = 3.74 R_Q$ (third less doped in Fig.\,\ref{resistance}, T = 8 K). The inset shows the second spectrum, or "`spectral wandering"' (obtained from the time dependence of the integrated voltage noise spectrum in the range $0.1$ Hz - $0.2$ Hz), which is found almost white (slope $-0.04$).}\label{noise}
\end{figure}

\begin{figure}
\resizebox{0.75\columnwidth}{!}{%
\includegraphics{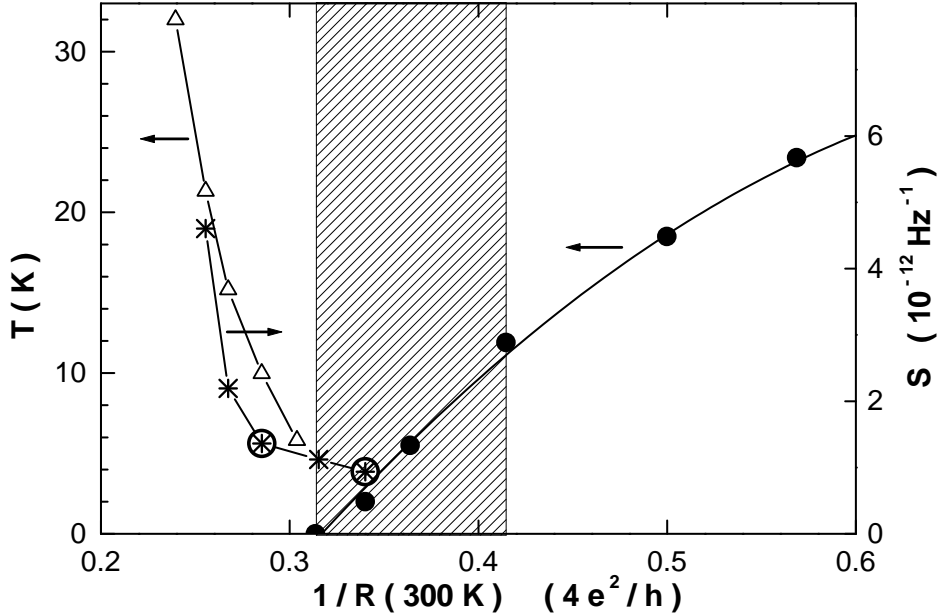}}
\caption{Superconducting transition temperature (full circles). The hatched area indicates the doping range where resistivity re-entrance is observed. Triangles is the VRH energy scale $T_0$ obtained from the construction in Fig.\,\ref{resistance}a. Stars represent the noise level at 8 K and 0.1 Hz, circled symbols correspond to the doping states of Fig.\,\ref{temperature} and to the 4$^{th}$ and $7^{th}$ less doped states in Fig.\,\ref{resistance}.}\label{diagram}
\end{figure}

In order to further characterize the interplay between the superconductor and the insulator, some complex decomposition of the transport data -- as was done in ref.~\cite{oh2006} -- is needed. Investigation of the noise transport properties does, however, provide an alternative way to gain information. Indeed, it is known that lightly doped semiconductors exhibit low--frequency resistance noise, determined by the charge distribution and hopping mechanism, as was observed in the vicinity of the superconductor-insulator transition for Bi$_2$Sr$_2$CaCu$_2$O$_{8+\delta}$\cite{fruchter2007}. In the present study, we also observed a power spectrum with frequency dependence close to $1/f$, as shown in Fig.\,\ref{noise}. No saturation at low frequency was found, and the noise was essentially Gaussian, as evidenced by the second spectrum analysis (Fig.\,\ref{noise}, inset), which indicates uncorrelated fluctuators and rules out glassy behavior (see Ref.~\cite{jaroszynski2002} and references therein). The noise level is found to rise steeply as one goes deep into the insulator, but a sizable level may still be measured at higher doping level, in the re-entrant regime (Fig.\ref{diagram}).

\begin{figure}
\resizebox{0.75\columnwidth}{!}{%
\includegraphics{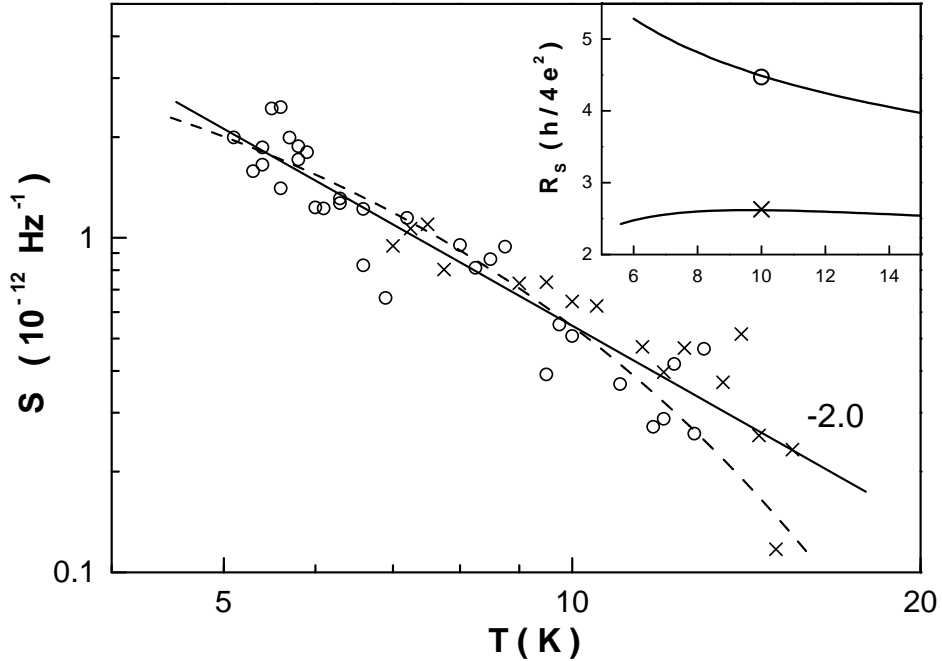}}
\caption{Noise spectrum temperature dependence for the sample in the VRH regime (circles) and in the re-entrant one (crosses) at 0.1 Hz. Line is best power law fit to the data and dashed line to exponential decay. The inset displays the corresponding resistivity data.}\label{temperature}
\end{figure}

We measured the temperature dependence of the noise, both in the re-entrant regime -- in a temperature range where superconductivity clearly manifests itself -- and at lower doping where the above analysis of the resistivity validated VRH mechanism. It may be seen in Fig.\,\ref{temperature} that the data, although scattered, is essentially similar for both regimes and may be well described by a power law, $S \propto T^{-2.0 \pm 0.1}$. We now compare this temperature dependence with the theoretical expectation for VRH low frequency noise, due to mobility fluctuations\cite{kozub1996,shklovskii2003}. 

According to the percolation theory of hopping charge transport in doped semiconductors\cite{shklovskii1984}, electrical conduction is intrinsically strongly inhomogeneoous at low temperature, due to the exponential dependence of hopping rates on randomly distributed distances between localized states. In the case of cuprates, holes, with a Bohr radius a few \AA\cite{edwards1995}, may be localized by the random distribution of the dopant (both Lanthanum and Oxygen in the present case). As a result, as described by Mott\cite{mott1968}, conduction in the insulator occurs along a critical resistor network with a typical bond length defined by the temperature-dependent hopping distance, $\xi(T)\propto T^{-1/3}$. Charges not making part of the critical network provide fluctuators that electrostatically modulate hopping along the critical path, which one assumes being 'two-sites fluctuators'. The resistance noise magnitude of the sample is then essentially determined by the critical network topology and, to a lesser extent, by the charge distribution in the vicinity of the critical network. The transposition of the three dimensional results of Ref.\cite{kozub1996} to the two dimensional one yields, for the VRH regime\cite{pokrovskii2001}:

\begin{equation}
S \approx N^{-1}(T) \frac{P_0 T}{\omega} D^2(T)
\nonumber
\label{pokrovskii}
\end{equation}  

where $N(T) \propto T^{-2(\nu+1)/(d+1)}$ is the number of fluctuating links of the resistive network ($\nu = 1.33$ in 2D ($d=2$)), $P_0$ is the density of two-sites charge fluctuators per unit energy (assumed exponentially broad) and $D^2(T)$ is a number of fluctuators. As shown in Ref.\cite{pokrovskii2001}, one expects $D(T) \propto T^{1/9}$ for the case of the dipolar fluctuator (electrostatic potential $\propto 1/r^3$; the distance between the charge sites is smaller than the distance to the critical link), and $D(T) \propto T^{-1/3}$, for the case where the Coulomb potential of one of the two sites dominates the hopping on the critical network (electrostatic potential $\propto 1/r^2$), yielding $S \propto T^{-1.44}$ and $S \propto T^{-1.89}$ respectively. Our data is in fair agreement with the latter case. There have also been more elaborate theoretical determinations of the noise magnitude in the VRH regime than this simple estimate. Shklovskii pointed out that remote fluctuators cannot not make part of the critical network\cite{shklovskii2003}, so that their number should decrease more strongly with increasing temperature (exponentially) than the simple power law used in Ref.~ \cite{kozub1996}. It was also proposed that many charges fluctuators can occur with much larger probability than single charge ones\cite{burin2006}. As a consequence, the simple power-law temperature decrease only holds when the fluctuators are randomly dispersed outside of the conducting planes, such that the total number of these sites does not vary with temperature. We cannot distinguish in the present case between a power law decrease and an exponential decrease, due to the restricted range of temperature of the study and scattering of the data (Fig.~\ref{temperature}). However, whatever the exact temperature dependence, a weak insulator is expected to exhibit negligible noise temperature dependence, considering the fact that we are far from a possible temperature crossover, $T_{CR}$, between the VRH and the NNH regime (we estimate $T/T_{CR} \simeq 0.3$ for the data in Fig.\ref{temperature}, yielding a few percent noise level variation, according to Ref.\cite{shklovskii2003}). 

At this point, we should stress that the agreement of the transport data with the results for the \textit{homogeneously} disordered insulator does not eliminate granularity, such as superconducting islands in an insulator matrix at doping close to the superconducting--insulator transition\cite{oh2006}, or nanoscale superconducting gap modulation for underdoped cuprates\cite{cren2000,kugler2001,davis2005,mashima2006}.
Indeed, it has been shown that granular systems may reproduce all the transport regimes obtained for homogeneously disordered materials. This is the case for VRH conductivity -- providing one includes the possibility for tunneling over long distances and many grains\cite{beloborodov2007} -- and, very likely also for noise properties. As a result, one cannot exclude that the `insulator' exhibiting VRH characteristics could be made from loosely coupled metallic grains, while the `metallic' islands could be made from strongly coupled ones.

Thus, the noise data indicates that the VRH insulator coexists in the present case with the superconductor, and that no weak insulating behavior may be found in the vicinity of the superconductor-insulator transition. This is in agreement with the data in Fig.\,\ref{diagram}, where it is seen that the VRH characteristic energy as a function of doping decreases as one approaches the re-entrant regime, but the slowing down of the decrease leaves open the possibility for a non--zero value deep into the shaded doping range, as is needed if the strong insulator coexists with the superconductor. (We note that there is no contradiction to assume VRH at temperatures larger than $T_0$, as long as the localized states energy distribution is large enough, Ref.~\cite{shklovskii1984}~\S 9). However, a clear $R \sim \log{(1/T)}$ behavior was observed in Ref.\cite{ono2000} for Bi$_2$Sr$_{2-x}$La$_x$CuO$_{6+\delta}$ ($x=0.84$). We note that this was found only once a magnetic field was applied, in a temperature range where superconductivity was otherwise clearly present at zero field. This, together with the present observations, suggest that the $\log{(1/T)}$ behavior -- if any -- is actually hidden by the superconductor and that superconductivity arises from metallic islands embedded in a strongly localized insulator, as evidenced by VRH hopping behavior. 

The observation that weakly insulating behavior is associated to doping states for which superconductivity occurs at sufficiently low temperature was early made in the case of hole-doped cuprates\cite{ando1996}. The $\log(1/T)$ behavior, in zero field and showing no saturation at low temperature, is observed only in the case of \textit{electron-doped} cuprates\cite{fournier2000}. This might be related to the fact that, for these coumpounds, superconductivity can be suppressed by the strong antiferromagnetism that overlaps the superconducting dome. A similar relationship between superconductivity, weak insulating behavior and the VRH insulator may be drawn in the case of the model system of ultrathin Bismuth films, either tuned by electrostatic field effect or magnetic field (Ref.\cite{parendo2006}). In this work, critical scaling -- attributed to a quantum superconductor-insulator transition -- was obtained only once a weak-localization contribution was subtracted from the metallic states conductivity: the insulating behavior that finally allowed for the critical scaling analysis was the one exhibiting VRH. In a general way, this suggests that the $\log(1/T)$ behavior is restricted to the metal that turns to a superconductor, at low temperature, in the absence of a pair breaker such as magnetic field or strong magnetic fluctuations; it is not intrinsic to the S-I transition, the strong insulator being the true phase associated to the S-I transition, embedding the metallic phase for heterogeneous materials, as in the present case.

\end{document}